\documentclass{emulateapj}

\input epsf.tex
\begin{document}

\shorttitle{The DA/DB WD Ratio}
\shortauthors{Kalirai et al.}

\title{The Dearth of Massive, Helium-rich White Dwarfs in Young Open Star Clusters\altaffilmark{1}}
\author{Jasonjot Singh Kalirai\altaffilmark{2}, Harvey B. Richer\altaffilmark{2}, Brad~M.~S.~Hansen\altaffilmark{3,4}, David Reitzel\altaffilmark{3}, \& R. Michael Rich\altaffilmark{3}}

\notetoeditor{If accepted, we would appreciate that this Letter and the following preceding 
submission entitled ``The Initial-Final Mass Relationship: Spectroscopy of White Dwarfs in 
NGC 2099 (M37)'' are published back to back, with this Letter coming second.}

\altaffiltext{1} {Based on observations with Gemini (run ID GN-2002B-Q-11) and Keck.
Gemini is an international partnership managed by the Association of
Universities for Research in Astronomy under a cooperative agreement with
the National Science Foundation.  The
W. M. Keck Observatory, which is operated as a scientific partnership
among the California Institute of Technology, the University of
California, and NASA, was made possible by the generous financial
support of the W. M. Keck Foundation.}
\altaffiltext{2}{Department of Physics and Astronomy, 6224 Agricultural Road, University of British 
Columbia, Vancouver, BC, V6T 1Z4, Canada; jkalirai@astro.ubc.ca, richer@astro.ubc.ca.}
\altaffiltext{3}{Department of Astronomy, University of California at Los Angeles, Box 951547, 
Knudsen Hall, Los Angeles, CA 90095-1547; hansen@astro.ucla.edu, reitzel@astro.ucla.edu, 
rmr@astro.ucla.edu.}
\altaffiltext{4}{Alfred P. Sloan Research Fellow.}

 \slugcomment{\it 
}

\lefthead{Kalirai, J. S. et al.}
\righthead{DA/DB White Dwarf Ratio}

\begin{abstract}

Spectra have been obtained of 21 white dwarfs (WDs) in the direction of the young,  rich 
open star cluster NGC 2099. This represents an appreciable fraction ($>30\%$) of the cluster's 
total WD population. The mean derived mass of the sample is 0.8~$M_{\odot}$ - about 0.2~$M_{\odot}$ 
larger than the mean seen among field WDs. A surprising result is that all of the NGC 2099 
WDs have hydrogen-rich atmospheres (DAs) and none exhibit helium-rich ones (DBs), or any other 
spectral class. The number ratio in the field at the temperatures of the NGC 2099 WDs is DA/DB 
$\sim 3.5$. While the probability of seeing no DB WDs in NGC 2099 solely by chance is $\sim 2\%$, 
if we include WDs in other open clusters of similar age it then becomes highly unlikely that the 
dearth of DB WDs in young open clusters is just a statistical fluctuation.  We explore possible 
reasons for the lack of DBs in these clusters and conclude that the most promising scenario for 
the DA/DB number ratio discrepancy in young clusters is that hot, high-mass WDs do not develop 
large enough helium convection zones to allow helium to be brought to the surface and turn a 
hydrogen-rich WD into a helium-rich one.
\end{abstract}

\keywords{open clusters and associations: individual (NGC 2099) - white dwarfs}

\section{Introduction}

It is by now well accepted that some white dwarfs (WDs) change their atmospheric chemical
composition as they cool. The fact that the ratio of DA to non-DA stars changes 
as a function of effective temperature renders this conclusion indisputable 
(see e.g., Bergeron, Ruiz, \& Leggett 1997).  The distribution of WDs from the 
Sloan Digital Sky Survey (SDSS, see Figure 11 of Kleinman et al. 2004) clearly shows 
that the overall fraction of hot DO WDs 
(those showing He {\sc ii} lines) first peaks and then declines at a temperature where the DB 
WDs (those showing He {\sc i} lines) begins to rise.  At cooler temperatures, the DC WDs 
(featureless spectrum) increase in number in the same temperature range that the DBs begin 
to decline (the DBs being too cool to show He {\sc i} spectral lines).  However, the 
distribution of WDs of various spectral types as a function of mass is much less constrained.  
Both the DA and DB WD mass distributions show a primary peak at $\sim$ 0.6 $M_\odot$ 
(Bergeron, Leggett, \& Ruiz 2001).  Very few, if any, massive DBs have been found with $M 
> 0.8 M_{\odot}$.

NGC 2099 is a very rich (4000 stars), young open star cluster in which we have found 50 WD candidates 
through a {\sl CFHT} imaging project (Kalirai et al. 2001a).  Given the age of this cluster 
(650 Myrs), most of these WDs have temperatures of 13,000~K--18,000~K.  At these temperatures, 
the number ratio of those exhibiting hydrogen-rich atmospheres (DAs) to those with helium 
atmospheres (DBs) is about 3.5/1 in the field (Kleinman et al. 2004).  In this 
{\it Letter}, we present spectroscopy of 21 WDs in NGC 2099 ($>$ 30\% of the entire cluster WD 
population). The effective temperatures are such that we might expect a number of DB white
dwarfs among their number, assuming the same statistics as in the field. Furthermore, the
contemporaneous nature of these stars may allow us to gather some insights into the poorly
understood physical mechanisms that underlie the DA/DB distinction.

\section{Observations}

Imaging and spectroscopic observations of NGC 2099 were obtained with the  
Canada-France-Hawaii ({\sl CFH}), {\sl Gemini North}, and {\sl Keck I} 
telescopes. In our wide field {\sl CFHT} imaging study (Kalirai et al.~2001a), 
we found 50 cluster WD candidates in the central 15$'$ 
of NGC 2099.  In the companion {\it Letter} (Kalirai et al. 2005), we describe the 
current observations and data reduction.  Summarizing, we obtained imaging and 
multi-object spectroscopy with both {\sc gmos} on {\sl Gemini} and {\sc lris} on {\sl Keck} of 
three small fields (5${.}^{\prime}$5 $\times$ 5${.}^{\prime}$5 for {\sl Gemini} 
and 5$' \times$ 7$'$ for {\sl Keck}).  These fields were chosen in an 
unbiased way to simply maximize the number of WD candidates for which we could 
obtain spectroscopy.  The {\sl Gemini} data have a high signal-to-noise ratio 
(22 1-hour exposures were obtained of the same field) whereas the {\sl Keck} data (2 fields 
were obtained) have both higher resolution and bluer spectral coverage, allowing 
the detection of higher order Balmer lines.  The observations of the second field 
were taken at high airmass and so the bluest flux has been lost as a result of 
atmospheric dispersion.  This, however, does not affect the classification of 
these stars as DA or DB. 

\begin{figure*}
\begin{center}
\leavevmode
\includegraphics[width=10.45cm,angle=270]{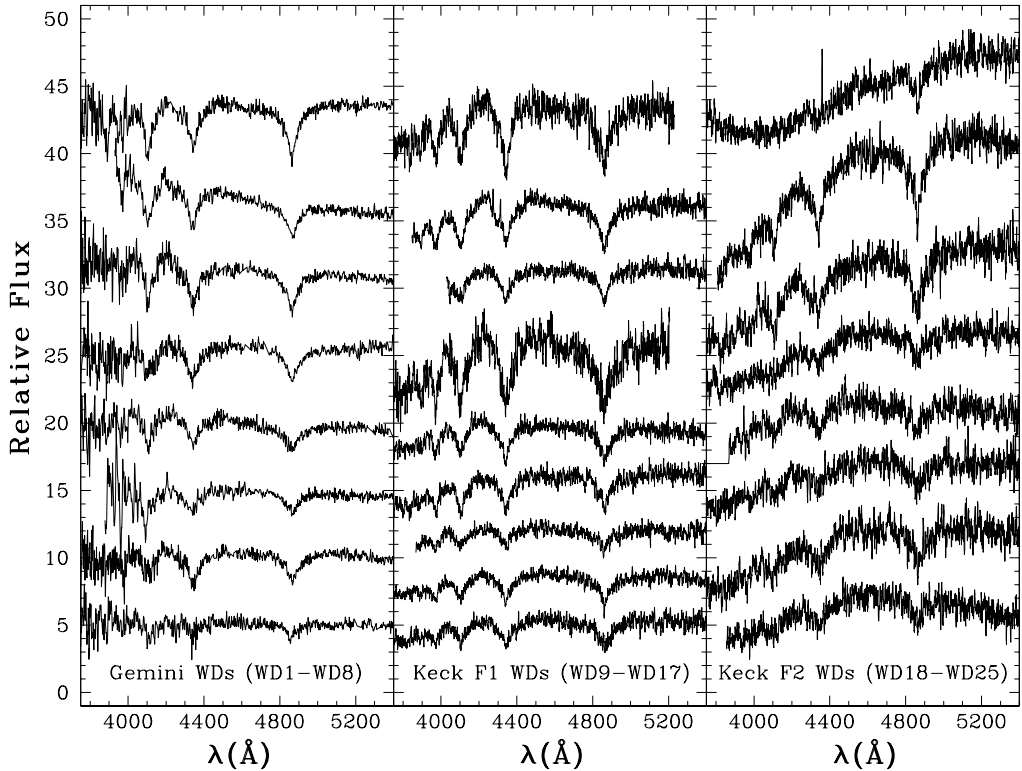}
\end{center}
\caption{Spectra of the 21 WDs observed in the direction of NGC 2099 shown in 
order of decreasing brightness from top (e.g., WD1) to bottom 
(e.g., WD8), within each panel (all with 22.46 $\leq V \leq$ 23.66).  Four of the spectra 
are shown twice: WD19 is the same star as WD1, WD20 is WD2, WD22 is WD5 and WD23 is WD6.  It 
is probable that 4-5 of the WDs are field objects.  Note that all of the WDs are DA spectral 
type.}
\label{spectra}
\end{figure*}

In total, we obtained spectroscopy of 24 individual WD candidates in the field 
of NGC 2099 (3 stars turned out not to be WDs).  Therefore, despite sampling only 14\% 
of the total cluster area, we include $>30\%$ of the total WD population (cluster 
and field) given the careful positioning of the fields.  This is therefore 
the largest individual star cluster WD sample that has ever been spectroscopically 
acquired.  We present the 25 spectra for the 21 WDs (4 stars are in common between the 
{\sl Gemini} and {\sl Keck} data) in Figure \ref{spectra}.  Surprisingly, all stars 
show hydrogen Balmer absorption lines and are of DA spectral type.  A table of the 
derived temperatures and masses for most of this sample of WDs is given in 
Kalirai et al. (2005).

\section{The Significance of our Sample}

Given our sample of 21 DA WDs, the 3.5/1 field DA/DB ratio at these temperatures suggests 
that we should see $\sim$ 4 DB WDs (considering that 4--5 of our 21 WDs are field objects).  
Assuming that binomial statistics are applicable, the probability of a sample producing 0 stars 
by chance with this expectation is $\sim1.4$\%. We can improve the statistics significantly by including 
spectroscopy from other young open clusters. The full sample in a search of the literature is 
shown in Table 1.  All of the turn-off stars in these clusters have masses above 
2~$M_{\odot}$ so that the WDs are all likely to be massive, just as in our sample.  The total number 
of hot, massive WDs observed rises to 61, where 13.5 are expected to be DBs, and none are found. 
The likelihood of this happening purely by chance is now 0.00002\%. 

Thus it appears as though massive open cluster WDs are always DAs, in contradiction
to the expectation from the overall field fraction. DBs are not normally 
found immediately above a temperature of 30,000~K (the DB-gap) nor are they found below a temperature 
of 12,000~K at which point helium lines become spectroscopically invisible (Bergeron, Ruiz, \& Leggett 
1997). (Of course, the stars may still possess helium atmospheres at these cool temperatures).  
All of our stars are between these temperatures, where DBs are abundant in the field. One possibility is 
that this indicates a hitherto unappreciated mass dependence to the phenomenon (\S~\ref{MassDep}).
Another is that the cluster environment may somehow lead to larger hydrogen envelope masses.

\begin{deluxetable}{lccl} 
\tablecolumns{4} 
\tablewidth{0pc} 
\tablecaption{White Dwarfs in Young Open Clusters \label{BigTab1}}
\tablehead{\colhead{Cluster}  & \colhead{\# DAs} & \colhead{\# DBs} & \colhead{Reference}}
\startdata 
Hyades & 6 & 0 & Eggen \& Greenstein (1965) \\
Pleiades & 1 & 0 & Luyten \& Herbig (1960) \\
Praesepe & 6 & 0 & Claver et al. (2001) \\
IC 2391 & 1 & 0 & Koester \& Reimers (1985) \\
NGC 1039 & 1 & 0 & Williams (2002) \\
NGC 2099 & 21 & 0 & This work (4-5 are field objects) \\
NGC 2168 & 8 & 0 & Williams, Bolte, \& Koester (2005)  \\
NGC 2287 & 2 & 0 & Koester \& Reimers (1981) \\
NGC 2422 & 1 & 0 & Koester \& Reimers (1981) \\
NGC 2451 & 3 & 0 & Koester \& Reimers (1985) \\
NGC 2516 & 4 & 0 & Reimers \& Koester (1982), (1996) \\
NGC 3532 & 6 & 0 & Reimers \& Koester (1989) \\ 
& & & Koester \& Reimers (1993) \\
NGC 6405 & 1 & 0 & Reimers \& Koester (1989) \\
NGC 6633 & 1 & 0 & Reimers \& Koester (1994) \\
NGC 6633 & 2 & 0 & Williams (2002) \\
NGC 7063 & 1 & 0 & Williams (2002) \\

\enddata 
\end{deluxetable} 

\section{The Role of the Cluster Environment}

Are cluster WDs really representative of 
field WDs?  The cluster environment is unlikely to give the WDs a common property 
(such as angular momentum, magnetic field, or rotation) as the initial conditions from the cloud 
collapse should be washed out by the huge disparity between cluster and stellar length scales 
(see e.g., Menard \& Duchene 2004).  White dwarfs in a given cluster will 
all have the same metallicity; however, our sample contains 15 clusters with different 
metallicities ranging from at least $Z$ = 0.008 to $Z$ = 0.024. Thus, it is reasonable to 
treat the 61 WDs as {\it individual} data points -- an assumption necessary for the 
statistical veracity of our result.  

The fact that star clusters are bound by their own gravity might lead one to imagine that 
they would trap gas lost in the advanced stages of stellar evolution and that this would 
lead to an enhanced accretion of gas (largely hydrogen) by the WDs in the cluster.  However, 
this notion is problematic from the outset. Open star clusters have low velocity dispersions
(NGC~2099 has $\sigma = 2.5$ km/s), much lower than the wind velocities from evolved 2--4 $M_{\odot}$ stars
($\sim 10$--30 km/s; e.g. Barnbaum, Zuckerman, \& Kastner 1991). As a result, the mass lost from the 
cluster is unbound and so the cluster WDs do not accrete from a gravitationally trapped reservoir 
of gas.  The WDs will, nevertheless, accrete some material from the escaping collective mass 
outflow. To determine whether cluster WDs might have thicker hydrogen layers than field 
WDs, we consider accretion from a `cluster wind'.

The wind velocity ($\sim$~20~km/s) is larger than the WD space velocity, so that the
 wind accretion rate 
($\dot{M} = 4 \pi \frac{G^2 M^2}{V^3} \rho$) onto a $0.85 M_{\odot}$ WD is
\begin{equation}
 \dot{M}  =  5.5 \times 10^{-16} M_{\odot}~yr^{-1}
\left( \frac{n}{1 cm^{-3}} \right).
\end{equation}
The real unknown here is $n$, the local gas density due to the outflowing wind.  Assuming 
the WD is located at a distance $r$ from the cluster center (which is outside the majority 
of the mass-losing stars), $n$ can be found from the cluster mass loss rate $\dot{M}_{\rm cl} 
= 4 \pi r^2 V m_{\rm p} n$, where $V$ is the wind velocity.  To get an expression for 
$\dot{M}_{\rm cl}$, we assume a Salpeter initial mass function for the cluster from 0.1~$M_{\odot}$ to 10~$M_{\odot}$.  
Since NGC 2099 is young enough that the remnant masses are small compared to the progenitor 
masses, we can also assume that all of the mass above the turnoff is lost.  Finally, to make 
this a function of time, we assume the turnoff mass scales with time as $T = 6~Gyr~( M/M_{\odot} 
)^{-2.5}$.  Putting this together, we arrive at the present day cluster mass loss rate, 
\begin{equation}
\dot{M_{\rm cl}} \sim 2.6 \times 10^{-7} M_{\odot}~yr^{-1} \left( \frac{T}{650 \, Myr} \right)^{-0.86}
\label{Mcleqn}
\end{equation}
\normalfont
where we have used $M_{\rm cl}$ = 2515~$M_{\odot}$ and turnoff mass $M$ = 2.4~$M_{\odot}$ 
(Kalirai et al. 2001a).

Equation~(\ref{Mcleqn}) can now be fed back to get $n$ and $\dot{M}$ for the WD as 
a function of cluster age.  This can then be integrated to yield the total 
amount of mass accreted by the WD, 
\begin{equation}
 M_{\rm acc} = 2.6 \times 10^{-8} M_{\odot} \left( \frac{r}{2 pc} \right)^{-2} \left(\frac{T}{650 \, Myr} \right)^{0.14}.
\label{Macceqn}
\end{equation}

The weak dependence on age is the result of the fact that the cluster mass loss rate
is higher at early times but lasts for a shorter period of time. Factors of 10 in age correspond
to about a 30\% variation in accreted mass. 

The important result to note here is that equation~(\ref{Macceqn}) does not lead to more 
accretion than would occur in the field. For a young field WD, moving at 20~km/s through an 
interstellar medium of density 1~$cm^{-3}$, $\dot{M} \sim 5 \times 10^{-16} M_{\odot}~yr^{-1}$. Thus, 
it appears unlikely that our observational result can be explained by an enhanced accretion rate 
for cluster WDs.  Thus we turn to the properties of the WDs themselves for an explanation.

\section{Is the DB phenomenon a function of White Dwarf mass?}
\label{MassDep}

If the incidence of non-DA stars is a strong function of WD mass, that may explain our data. 
In fact, the results of Beauchamp et al. (1996) contain a hint to this effect. They find that
the mass distribution of DB stars lacks the tail to high masses that is seen in the DA mass
distribution (and even in the smaller DBA sample). Our result makes this possibility much
stronger statistically.  What are the possible physical mechanisms for such a mass dependence?

\subsection{The Mass Dependence of Convective Mixing}

Although the chemical evolution of WDs is well known, the exact manner in which 
these transformations are achieved is still a matter of discussion.  The appearance of DB 
stars at temperatures less than 30,000~K is thought to be due to the onset of convection 
in the helium layer that lies beneath the surface hydrogen layer (Liebert, Fontaine, 
\& Wesemael 1987). If the hydrogen surface layer is thin enough (Liebert et al. considered 
very thin layers $\sim 10^{-14} M_{\odot}$) then it spans only a few pressure scale 
heights and convective overshooting may overwhelm a surface hydrogen layer.  The status 
of this idea in the face of later estimates of surface hydrogen layer masses 
(Clemens 1993) is unclear.

Although WD models do not yet exist that contain a detailed treatment of convective
overshooting, we can at least map the mass dependence of the onset of convection in the
helium layer for a variety of hydrogen surface layer masses, using the models of
Hansen (1999).  For $q(\rm H)$ = 
$M_{\rm H}/M$ = $10^{-7}$ or smaller, WDs with masses $\lesssim$~1~$M_\odot$ will 
form convective zones in the subsurface helium layer (provided the WDs are cooler than 
$\sim$~25,000~K).  However, as the surface hydrogen mass layer becomes larger, there is 
an upper mass limit (near 0.8 $M_\odot$ for $q(\rm H) = 10^{-6}$) above which 
a convective zone will never form in the helium layer.  Therefore, given the higher masses 
of the NGC 2099 WDs (mostly 0.7--0.9 $M_\odot$ - see Table 1 of Kalirai et al. 2005) 
compared to field WDs, the conversion of a DA to a DB WD through 
convective mixing may be inhibited by hydrogen layer masses that would still
allow the transformation of 0.6~$M_{\odot}$ WDs.
 Note, however, that these layers are considerably thicker than in the 
original proposal of Liebert et al. (1987), and we are unaware of any quantitative model that 
has yet demonstrated the kind of mixing required in models with these parameters.

\newpage

\subsection{Mass Dependence of Hydrogen Removal}

It is also possible that the mechanisms that determine the final hydrogen surface mass are
mass dependent as well. The most likely value of $q(\rm H)$ from stellar evolution is $q(\rm H) 
\sim 10^{-4}$ (Iben \& Tutukov 1984), a value high enough that it is likely to resist any 
attempt at convective mixing. The mechanisms 
that produce lower hydrogen masses are poorly understood but some are indeed likely to depend on
mass.  For example, Lawlor \& MacDonald (2003) show that ``born-again'' stars will be devoid 
of hydrogen after leaving the asymptotic giant branch for the final time.  These stars evolve 
through this phase as a result of changes in the efficiency of convective mixing, which may be mass 
dependent.  Other mechanisms, such as the `self-induced nova' of Iben \& MacDonald (1986) rely on 
shell flashes powered by CNO burning and so may be more easily quenched in more massive WDs where the 
heavier elements separate out more rapidly in the higher gravity. The competition between nuclear
burning and gravitational separation in massive WDs is another issue that needs to be investigated
anew in the light of these results.
 
\subsection{Binary Evolution}

Finally, it is possible that a number of WDs in NGC 2099 have been affected by some 
type of binary star evolution (interacting, merging, etc.).  There is evidence to suggest 
binary evolution results in fewer DB white dwarfs (Liebert, Bergeron, \& Holberg 2004, \S~5.2), and 
this could alter the cluster DA/DB ratio (although binaries with non-DA components do exist, 
e.g., Wood \& Oswalt 1992).  Hurley \& Shara (2003) have recently modelled the WD cooling 
sequence of NGC 6819 (see Kalirai et al. 2001b) using $N$ body simulations, 
and they found that the observed morphology of the cooling sequence can be understood by invoking a large double 
degenerate and/or mass transfer WD binary population.  These simulations include contributions from a 
number of different binary evolutionary scenarios, such as common envelope evolution and mass transfer 
systems.  Such effects could also play a role in the evolution of NGC 2099 WDs, which do not exhibit 
a tight sequence of points on the color-magnitude diagram but rather a more diffuse clump (see 
Kalirai et al. 2001a).  For a 40\% primordial binary fraction, Hurley \& Shara (2003) estimate 
that 12\% of all WDs will be double degenerates after 0.5 Gyrs (J. Hurley, 2004, private 
communication).  However, up to 39\% of the WDs are nonstandard single WDs (i.e., they were 
in binaries at some point and may have suffered mass exchange) in which evolution may have 
resulted in mass transfer and thick H envelopes. 

\section{Conclusions}

Spectroscopic observations in a sample of young open star clusters have shown that all cluster WDs 
exhibit hydrogen-rich atmospheres.  This is in contrast to WDs in the field, where at temperatures between 
13,000~K and 18,000~K, $\sim$ 25\% are of DB type. The statistics are now good enough that the probability 
of this happening purely by chance is vanishingly small.  A potential explanation is that the cluster 
environment might be the culprit. However, the only likely physical mechanism here -- the accretion of
H-rich gas produced by massive evolving stars -- is found to be not higher than the rate for field objects. 
The most promising scenario appears to be related to the fact that these cluster WDs are, on average, 
more massive than those in the field because of the young ages of their parent clusters.  Cooling models 
(Hansen 1999) suggest that the underlying helium convection zone may not easily break through to the surface 
hydrogen layer in massive WDs as it does for average mass WDs.  This scenario makes two clear 
testable predictions: (1) field DB WDs should be less massive on average than field DAs at temperatures 
below about 20,000~K, and (2) the field DA/DB ratio ought to be found in older clusters whose 
turnoff stars produce WDs nearer to the canonical $0.6 M_{\odot}$.
 
\acknowledgements
We would like to thank Pierre Bergeron for providing us with his
models and spectral fitting routines.  This project could not have
succeeded without this input.  We also wish to thank Dan Kelson, Inger
Jorgenson, and Jarrod Hurley for providing help with various parts of this
project.  J.~S.~K. received financial support during this work through an
NSERC PGS-B research grant.  The research of H.~B.~R. is supported in part
by NSERC.  B.~M.~S.~H. is supported by NASA grant ATP03-0000-0084.  B.~K.~G.
acknowledges the financial support of the Australian Research Council.
T.~v.~H. appreciatively acknowledges support from NASA through LTSA grant
NAG5-13070.

\end{document}